\documentclass[11pt]{article}

\usepackage{amsmath}

\begin{document}

\title{
 V.M. Red'kov,  E.M.   Ovsiyuk, V.V. Kisel\\[5mm]
 Spin 1 particle on
4-dimensional sphere: extended helicity operator, separation of
the variables, and exact solutions\\[2mm]
{\small redkov@dragon.bas-net.by; e.ovsiyuk@mail.ru}
}

\date{}
\maketitle

\begin{abstract}
Spin 1  particle is investigated in  3-dimensional curved
space of constant positive curvature. An extended helicity
operator is defined and the variables are separated in a
tetrad-based 10-dimensional Duffin-Kemmer equation in quasi
cylindrical coordinates. The problem is solved exactly in
hypergeometric functions, the energy spectrum determined by three
discrete quantum numbers is obtained. Transition to a massless
case of electromagnetic field is performed.

\end{abstract}

In 3-dimensional spherical Riemann space $S_{3}$  will use the following system of quasi-cylindric coordinates
(see \cite{Olevsky-1950}; the same coordinate system was used when treating
Landau problem in 3-dimensional spaces of constant curvature  in \cite{Bogush-Red'kov-Krylov-2008, Bogush-Red'kov-Krylov-2009,
Kisel-Ovsiyuk-Veko-Red'kov-2012})
\begin{eqnarray}
 dS^{2} = c^{2} dt^{2} - \rho^{2}
\; [ \; \cos^{2} z ( d r^{2} + \sin^{2} r \; d \phi^{2} ) +
dz^{2}\; ]\; ,
\nonumber
\\
z \in [-\pi /2 , + \pi /2 ]\; , \qquad r \in [0, + \pi ] , \qquad
\phi \in [0, 2 \pi ] \; ;
\label{4.1}
\end{eqnarray}

\noindent a diagonal tetrad (let  $x^{\alpha}= (t,r,\phi,z)$)
\begin{eqnarray}
 e_{(a)}^{\beta}(x) = \left |
\begin{array}{llll}
1 & 0 & 0 & 0 \\
0 & \cos^{-1}z & 0 & 0 \\
0 & 0 & \cos^{-1}z\;\sin^{-1} r & 0 \\
0 & 0 & 0 & 1
\end{array} \right | \; ;
\label{4.2}
\end{eqnarray}

 \noindent corresponding
 Christofel ant Ricci coefficients are
\begin{eqnarray}
\Gamma^{r}_{\;\;jk } = \left | \begin{array}{ccc}
0 & 0 & -\tan z \\
0 & - \sin r \cos r & 0 \\
- \tan z & 0 & 0
\end{array} \right | \; ,
\nonumber
\\
\Gamma^{\phi}_{\;\;jk } = \left | \begin{array}{ccc}
0 & \cot r & 0\\
\cot  r & 0 &- \tan  z \\
0 & -\tan  z & 0
\end{array} \right | \; ,
\nonumber
\\
\Gamma^{z}_{\;\;jk } = \left | \begin{array}{ccc}
\sin z \cos z & 0 & 0\\
0 & \sin z \; \cos z \sin^{2} r & 0 \\
0 & 0 & 0
\end{array} \right | \; ,
\nonumber
\\
\gamma_{12 2} =
 { 1 \over \cos z \tan r} \; , \qquad
 \gamma_{31 1} =
 -\tan z\; , \qquad \gamma_{32 2} =
 -\tan  z\; .
 \label{4.3}
 \end{eqnarray}

Tetrad-based  Duffin-Kemmer  equation (the notation from \cite{book-2009}  is used)  takes the form
\begin{eqnarray}
\left \{  i \beta^{0} { \partial \over \partial t} + {1 \over \cos
z} \left (  i  \beta^{1} { \partial \over \partial r}        +
\beta^{2}      { i  \partial _{ \phi}  + i J^{12} \cos r  \over
\sin r}  \right )   \right.
\nonumber
\\
 \left.
  +
   i\beta^{3}  { \partial \over \partial  z}  +  i   {\sin z \over \cos z}  \; ( \beta^{1} J^{13}    +   \beta^{2} J^{23} )
          - M   \right \} \Psi  = 0 \; .
\label{4.4}
\end{eqnarray}

To separate the variables, we take the substitution
\begin{eqnarray}
\Psi = e^{-i\epsilon t  }  e^{im\phi}   \left |
\begin{array}{c}
\Phi_{0}  (r,z) \\
\vec{\Phi}(r,z)  \\
\vec{E} (r,z)  \\
\vec{H} (r,z)
\end{array} \right | 
\label{4.5a}
\end{eqnarray}

\noindent and use a block-representation (we use so-called cyclic basis for $10 \times 10$ Duffin-Kemmer matrices -- see in
\cite{book-2011})
\begin{eqnarray}
 \left  [ \;  \epsilon \; \cos z\;   \left | \begin{array}{rrrr}
 0       &   0        &  0  &  0 \\
 0  &  0       &  i  & 0  \\
  0  &   -i       &   0  & 0\\
   0  &  0       &   0  & 0
\end{array} \right |    + i\;  \left |
\begin{array}{rrrrr}
  0       &  0       &    e_{1}  & 0       \\
    0   &  0       &   0      & \tau_{1} \\
   -e_{1}^{+}  &  0       &   0      & 0       \\
   0       &  -\tau_{1}&   0      & 0
\end {array} \right | {\partial \over \partial r }  \right.
\nonumber
\\
 - {1 \over  \sin r  } \; \left |
\begin{array}{rrrrr}
  0       &  0       &    e_{2}  & 0       \\
    0   &  0       &   0      & \tau_{2} \\
   -e_{2}^{+}  &  0       &   0      & 0       \\
   0       &  -\tau_{2}&   0      & 0
\end {array} \right |
 (   \nu     - \cos  r \; S_{3})
\nonumber
\\
   + i \cos z\;   \left |
\begin{array}{rrrrr}
  0       &  0       &    e_{3}  & 0       \\
    0   &  0       &   0      & \tau_{3} \\
   -e_{3}^{+}  &  0       &   0      & 0       \\
   0       &  -\tau_{3}&   0      & 0
\end {array} \right | {\partial \over \partial z}
\nonumber
\\
\left. + i \sin z \left | \begin{array}{cccc}
0 & 0 &  -2 e_{3}  & 0 \\
0 & 0 & 0 &  - \tau_{3}  \\
0 & 0 &  0   & 0 \\
0 & + \tau_{3} & 0 & 0
\end{array} \right |
  - M\; \cos z \;  \right  ]  \left | \begin{array}{c}
\Phi_{0} \\
\vec{\Phi} \\
\vec{E} \\
\vec{H}
\end{array} \right |
 = 0 \;  ,
\label{4.5b}
\end{eqnarray}

\noindent or
\begin{eqnarray}
   i e_{1}  \partial _{r } \vec{E}
   -  {1 \over  \sin r} \; e_{2} (m  - \cos r \,s_{3} )  \vec{E}
    +
    \nonumber
    \\
    +  i  (\cos z\;  \;  \partial _{z }  - 2 \sin z ) e_{3}\vec{E}= M \cos z   \; \Phi_{0} \; ,
\nonumber
\\
i \epsilon \cos z\, \vec{E} + i \tau_{1} \partial_{r} \vec{H} -
{\tau_{2} \over \sin r } ( \;  m  - \cos r \, s_{3} \;) \vec{H}  +
\nonumber
\\
+
i  ( \cos z\; \partial _{z }   - \sin z ) \tau_{3}\vec{H} =M \cos
z  \; \vec{\Phi} \; ,
\nonumber
\\
-i\epsilon \cos z\, \vec{\Phi}  -i e_{1}^{+} \partial_{r}
\Phi_{0} + {m \over  \sin r} \,e_{2}^{+} \Phi_{0}  - i \cos z\;
e_{3}^{+}\partial _{z } \Phi_{0} = M \cos z \;  \vec{E}\; ,
\nonumber
\\
-i \tau_{1} \partial_{r} \vec{\Phi} + { (m -\cos r \, s_{3}) \over
\sin r} \,\tau_{2}  \vec{\Phi}  - i ( \cos z\; \partial _{z } -
\nonumber
\\
-
i\sin z ) \tau_{3}\vec{\Phi}= M \cos z \; \vec{H}\; .
\label{4.5c}
\end{eqnarray}

\noindent After calculations needed we arrive at the system
\begin{eqnarray}
  \gamma  (  { \partial  E_{1} \over \partial  r} - {\partial E_{3} \over \partial r
})  -  {\gamma \over  \sin r }   \left [  (m -\cos r ) E_{1} + (m
+\cos r) E_{3}   \right ]   -
\nonumber
\\
-(  \cos z {\partial  \over \partial z} - 2 \sin z ) E_{2}
 = M \cos z   \Phi_{0} \; ,
\label{4.6a}
\end{eqnarray}
\begin{eqnarray}
+i \epsilon \cos z  E_{1}    +   i \gamma  {\partial H_{2} \over
\partial  r}
 +   i\gamma     { m \over  \sin r }   H_{2}
 +
 \nonumber
\\
+  i  ( \cos z   {\partial   \over \partial  z}- \sin z ) H_{1}= M \cos z \Phi_{1}\;,
\nonumber
\\
+i \epsilon  \cos z  E_{2}  +  i \gamma ( {\partial  H_{1} \over
\partial  r } + {\partial  H_{3} \over \partial  r } ) -
\nonumber
\\
-
 {i\gamma
\over  \sin r } \left [  (m -\cos r)  H_{1}  -   (m + \cos r)
H_{3}  \right ]  = M \cos z  \Phi_{2}\;,
\nonumber
\\
+i \epsilon  \cos z  E_{3}    +  i \gamma\; {\partial   H_{2}
\over \partial r}  -   i \gamma  {m  \over  \sin r }  H_{2} -
\nonumber
\\
- i
(\cos z {\partial  \over \partial z} - \sin z ) H_{3} = M \cos z
\Phi_{3}\,,
\label{4.6b}
\end{eqnarray}
\begin{eqnarray}
-i  \epsilon  \cos z \Phi_{1}  +  \gamma   {\partial \Phi_{0}
\over d r}   +
 \gamma  {m \over  \sin r }   \Phi_{0} = M \cos z   E_{1}\;,
\nonumber
\\
-i \epsilon   \Phi_{2}  -   {\partial  \Phi_{0} \over \partial  z}
= M E_{2}\;,
\nonumber
\\
-i \epsilon \cos z \Phi_{3}  -   \gamma   {\partial  \Phi_{0}
\over \partial r}  +
 \gamma {m   \over  \sin r }   \Phi_{0} = M \cos z  E_{3}\;,
\label{4.6c}
\end{eqnarray}
\begin{eqnarray}
-i  \gamma   {\partial  \Phi_{2} \over \partial  r}  -  i \gamma
{m   \over  \sin r }  \Phi_{2}- i ( \cos z{\partial  \over
\partial  z} - \sin z ) \Phi_{1} = M \cos z  H_{1}\;,
\nonumber
\\
 - i  \gamma  (  {\partial  \Phi_{1} \over \partial r } + { \partial \Phi_{3} \over \partial r })
+  {i \gamma \over  \sin r }  [ (m -\cos r) \Phi_{1} - (m+\cos
r)\Phi_{3}  ]  = M  \cos z  H_{2} \; ,
\nonumber
\\
-i  \gamma   {\partial  \Phi_{2} \over \partial r}  +
 i \gamma {m \over  \sin r }   \Phi_{2}
  +i (  \cos z  { \partial   \over \partial  z}-  \sin z ) \Phi_{3}=  M \cos z   H_{3}\;.
\nonumber
\\
\label{4.6d}
\end{eqnarray}

With the use of the notation
\begin{eqnarray}
\gamma \;  ( {\partial  \over \partial r} +  { m -\cos r  \over
\sin r } ) = a_{-} , \nonumber
\\
 \gamma \;  ( {\partial  \over \partial
r} +  { m +\cos r  \over  \sin r } ) = a_{+}\;, \nonumber
\\
 \gamma \;
( {\partial  \over \partial r} +  { m   \over  \sin r } ) = a \; ,
\nonumber
\\
\gamma \;  (- {\partial  \over \partial r} +  { m -\cos r  \over
\sin r } ) = b_{-} ,
\nonumber
\\
 \gamma \;  (- {\partial  \over
\partial r} +  { m +\cos r  \over  \sin r } ) = b_{+}\;,
\nonumber
\\
\gamma \;  (- {\partial  \over \partial r} +  { m   \over  \sin r
} ) = b \; ,
\label{4.7}
\end{eqnarray}

\noindent it  reads simpler
\begin{eqnarray}
  -b_{-}  \; E_{1}   - a_{+}  \;  E_{3}
-   \cos z ( {\partial  \over
\partial z } -2 \tan z) E_{2}
 = M   \; \cos z \Phi_{0} \; ,
\label{4.8a}
\end{eqnarray}
\begin{eqnarray}
        i   a  \;    H_{2} +i \epsilon  \; \cos z   E_{1}
 + i  \cos z ({\partial     \over \partial z}- \tan z) H_{1}  = M  \;  \cos z  \Phi _{1}\;,
\nonumber
\\
   - i  b_{-}  \;  H_{1}    +  i a_{+} \; H_{3}
    + i \epsilon \; \cos z     E_{2}   = M  \; \cos z   \Phi_{2}\;,
\nonumber
\\
  - i  b\;  H_{2} +i \epsilon  \;\cos z    E_{3}
- i ( {\partial     \over \partial z} - \tan z) H_{3}  = M    \;
\cos z\; \Phi_{3}\; , \label{4.8b}
\end{eqnarray}
\begin{eqnarray}
    a \; \Phi_{0}  - i  \epsilon  \; \cos z  \Phi_{1} = M \; \cos z     E_{1}\;,
\nonumber
\\
-i \epsilon    \Phi_{2}   -   {\partial  \over \partial z}  \Phi
_{0} = M  \;   E_{2} \;,
\nonumber
\\
    b \;
\Phi_{0}  -i \epsilon  \;\cos z \Phi _{3}  = M    \; \cos z
E_{3}\;, \label{4.8c}
\end{eqnarray}
\begin{eqnarray}
-  i  a\;  \Phi_{2} \;- i  \cos z ({\partial
 \over \partial z}  -\tan z) \Phi_{1} = M  \;\cos z  H_{1}\;,
\nonumber
\\
  i   \; b_{-}  \; \Phi_{1}
 - i  a_{+} \; \Phi_{3}     = M  \;  \cos z  H_{2} \; ,
\nonumber
\\
  i  \; b  \;  \Phi_{2}
  + i  \cos ({\partial    \over \partial z} - \tan z )  \Phi_{3} =  M  \;  \cos   H_{3}\;.
\label{4.8d}
\end{eqnarray}

Let us employ  additional operator, a generalized helicity operator -- such that
\begin{eqnarray}
 \Sigma \Psi = \sigma \Psi,\qquad
 \Psi = e^{-i\epsilon t  }  e^{im\phi}   \left |
\begin{array}{c}
\Phi_{0}  (r,z) \\
\vec{\Phi}(r,z)  \\
\vec{E} (r,z)  \\
\vec{H} (r,z)
\end{array} \right | ,
\nonumber
\end{eqnarray}
\begin{eqnarray}
\left [ {1 \over \cos z} \left ( S_{1} \; {\partial  \over
\partial r }  + iS_{2}
 { m - S_{3} \cos r  \over \sin r }\right )
+  ( {\partial \over \partial z} -\tan z)
 \;S_{3}
  \right ] \left |
\begin{array}{c}
\Phi_{0}   \\
\vec{\Phi}  \\
\vec{E}   \\
\vec{H}
\end{array} \right |= \sigma \;   \left |
\begin{array}{c}
\Phi_{0}   \\
\vec{\Phi}  \\
\vec{E}   \\
\vec{H}
\end{array} \right |\,.
\label{4.9}
\end{eqnarray}

\noindent From (\ref{4.9}) it follows the system of 10 equations
(let $\gamma =1 /
\sqrt{2}$):
\begin{eqnarray}
0= \sigma \; \Phi_{0}\,, \label{4.10a}
\end{eqnarray}
\begin{eqnarray}
  \gamma   {\partial  \over \partial  r }   \Phi_{2}  +\gamma   { m   \over   \sin r }  \Phi_{2}
+ \cos z ({ \partial  \over \partial  z} -\tan z )  \Phi_{1} =
\sigma\, \cos z\;  \Phi_{1} \;,
\nonumber
\\
  \gamma  (  {\partial  \over \partial  r }   \Phi_{1}  + {\partial  \over \partial  r } \Phi_{3} )
-
\nonumber
\\
-{\gamma \over  \sin r }  [ (m - \cos r) \Phi_{1} - (m + \cos r
)\Phi_{3}  ]  = \sigma  \,  \cos z  \; \Phi_{2} \; ,
\nonumber
\\
  \gamma   {\partial  \over \partial  r }   \Phi_{2}  -
  \gamma {m  \over   \sin r }   \Phi_{2}
  - \cos z ({ \partial  \over \partial  z} -\tan z )   \Phi_{3}=  \sigma \;  \cos z   \; \Phi_{3}\;,
\label{4.10b}
\end{eqnarray}
\begin{eqnarray}
  \gamma   {\partial  \over \partial  r }   E_{2}  +\gamma   { m   \over   \sin r }  E_{2}
+ \cos z ({ \partial  \over \partial  z} -\tan z )  E_{1} =
\sigma\, \cos z  \; E_{1} \;,
\nonumber
\\
  \gamma  (  {\partial  \over \partial  r }   E_{1}  + {\partial  \over \partial  r } E_{3} )
-
\nonumber
\\
-{\gamma \over  \sin r }  [ (m - \cos r) E_{1} - (m + \cos r
)E_{3}  ]  = \sigma  \,  \cos z  \; E_{2} \; ,
\nonumber
\\
  \gamma   {\partial  \over \partial  r }   E_{2}  -
  \gamma {m  \over   \sin r }   E_{2}
  - \cos z ({ \partial  \over \partial  z} -\tan z )  E_{3}=  \sigma \;  \cos z \;  E_{3}\;,
\label{4.10c}
\end{eqnarray}
\begin{eqnarray}
  \gamma   {\partial  \over \partial  r }   H_{2}  +\gamma   { m   \over   \sin r }  H_{2}
+ \cos z ({ \partial  \over \partial  z} -\tan z )  H_{1} =
\sigma\, \cos z\; H_{1} \;,
\nonumber
\\
  \gamma  (  {\partial  \over \partial  r }   H_{1}  + {\partial  \over \partial \sin r } H_{3} )
-
\nonumber
\\
- {\gamma \over  \sin r }  [ (m - \cos r) H_{1} - (m + \cos r
)H_{3}  ]  = \sigma  \,  \cos z  \;H_{2} \; ,
\nonumber
\\
  \gamma   {\partial  \over \partial  r }   H_{2}  -
  \gamma {m  \over   \sin r }   H_{2}
  - \cos z ({ \partial  \over \partial  z} -\tan  z )  H_{3}=  \sigma \;  \cos z   \;H_{3}\;.
\label{4.10d}
\end{eqnarray}

\noindent With notation (\ref{4.7}) it reads simpler
\begin{eqnarray}
0= \sigma \; \Phi_{0}\,,
\label{4.11a}
\end{eqnarray}
\begin{eqnarray}
  + \cos z ({ \partial  \over \partial  z} -\tan z )  \Phi_{1} = \sigma\, \cos z \Phi_{1} - a \Phi_{2}  \;,
\nonumber
\\
- b_{-} \Phi_{1} + a_{+} \Phi_{3} = \sigma \; \cos z \; \Phi_{2}
  \; ,
\nonumber
\\
   - \cos z ({ \partial  \over \partial  z} -\tan z )   \Phi_{3}=  \sigma \;  \cos z   \Phi_{3} +  b\;   \Phi_{2} \;,
\label{4.11b}
\end{eqnarray}
\begin{eqnarray}
 + \cos z ({ \partial  \over \partial  z} -\tan z )  E_{1} = \sigma \; \cos z \; E_{1} -a\;    E_{2}  \;,
\nonumber
\\
- b_{-} E_{1} + a_{+} E_{3} = \sigma \; \cos z \;  E_{2}
  \; ,
\nonumber
\\
     - \cos z ({ \partial  \over \partial  z} -\tan z )   E_{3}=  \sigma \;  \cos z  \; E_{3} + b\;    E_{2}\;,
\label{4.11c}
\end{eqnarray}
\begin{eqnarray}
   + \cos z ({ \partial  \over \partial  z} -\tan z )  H_{1} = \sigma\, \cos z H_{1} -  a\;    H_{2}\;,
\nonumber
\\
- b_{-} H_{1} + a_{+} H_{3} = \sigma\; \cos z \;  H_{2}
  \; ,
\nonumber
\\
    - \cos z ({ \partial  \over \partial  z} -\tan z )    H_{3}=  \sigma \;  \cos z  \; H_{3} + b\;    H_{2} \;.
\label{4.11d}
\end{eqnarray}

Taking into account  eqs.  (\ref{4.11a}) -- (\ref{4.11d}), from  (\ref{4.8a}) -- (\ref{4.8d}) we get
\begin{eqnarray}
  -b_{-}  \; E_{1}   - a_{+}  \;  E_{3}
-   \cos z ( {\partial  \over
\partial z } -2 \tan z) E_{2}
 = M   \; \cos z \Phi_{0} \; ,
\label{4.12a}
\end{eqnarray}

\begin{eqnarray}
        i \epsilon  \;  E_{1}
 + i  \sigma\,  H_{1}  = M  \;  \Phi _{1}\;,
\nonumber
\\
  i \sigma\;  H_{2}
    + i \epsilon \;    E_{2}   = M  \; \Phi_{2}\;,
\nonumber
\\
  i \epsilon  \;  E_{3}
+ i \sigma \;  H_{3}   = M    \; \Phi_{3}\; , \label{4.12b}
\end{eqnarray}
\begin{eqnarray}
    a \; \Phi_{0}  - i  \epsilon  \; \cos z  \Phi_{1} = M \; \cos z     E_{1}\;,
\nonumber
\\
-i \epsilon    \Phi_{2}   -  {\partial  \over \partial z}  \Phi
_{0} = M  \;   E_{2} \;,
\nonumber
\\
    b \;
\Phi_{0}  -i \epsilon  \;\cos z \Phi _{3}  = M    \; \cos z
E_{3}\;, \label{4.12c}
\end{eqnarray}
\begin{eqnarray}
-   \sigma\,  \Phi_{1}   = M  \; H_{1}\;,
\nonumber
\\
  - i \sigma \;  \Phi_{2}    = M  \;   H_{2} \; ,
\nonumber
\\
 -i\sigma \;    \Phi_{3}   =  M  \;    H_{3}\;.
\label{4.12d}
\end{eqnarray}

Below we will need an explicit form of the Lorentz condition. Starting from its tensor form
\begin{eqnarray}
 \nabla_{\beta} ( e^{(b)\beta} \Phi_{(b)}^{cart}  =0\qquad
\Longrightarrow
\nonumber
\\
 { \partial \Phi_{(b)^{cart} } \over \partial x^{\beta} }
   \;    e^{(b)\beta} +   \Phi_{(b)} ^{cart}  \nabla_{\beta}  e^{(b) \beta}   =0 \; ,
\label{4.13a}
\end{eqnarray}

\noindent or
\begin{eqnarray}
 { \partial \Phi_{(b) }^{cart} \over \partial x^{\beta} }
   \;    e^{(b)\beta} +   \Phi_{(b)}^{cart}  { 1 \over \sqrt{-g}} {\partial \over \partial x^{\beta}}
\sqrt{-g}   e^{(b) \beta}   =0 \; , \label{4.13b}
\end{eqnarray}

\noindent and taking into consideration (\ref{4.1})--(\ref{4.2}), we transform eq. (\ref{4.13b}) to the form
\begin{eqnarray}
{\partial \over \partial t} \Phi_{0}^{cart} -
 {1 \over \cos z} {\partial \over \partial r} \Phi^{cart}_{1} -
{1 \over \cos z \sin r} {\partial \over \partial \phi }
\Phi^{cart}_{2} - {\partial \over \partial z} \Phi^{cart}_{3} -
\nonumber
\\
-\Phi^{cart}_{1} {1 \over \cos^{2} z \sin r} {\partial \over
\partial r} \cos^{2} z \sin r \; {1 \over  \cos z} -
 \Phi^{cart}_{3} {1 \over \cos^{2} z \sin r} {\partial \over \partial z} \cos^{2} z \sin r \; =0 \; ,
\nonumber
\end{eqnarray}

\noindent that is
\begin{eqnarray}
{\partial \over \partial t} \Phi_{0}^{cart} -
 {1 \over \cos z} ({\partial \over \partial r} +{ \cos r \over  \sin r}) \Phi^{cart}_{1} -
 \nonumber
\\
-
{1 \over \cos z \sin r} {\partial \over \partial \phi }
\Phi^{cart}_{2} - ( {\partial \over \partial z} - 2 \tan z)
\Phi^{cart}_{3} =0 \; .
\nonumber
\end{eqnarray}

\noindent From whence with the substitution (\ref{4.5a})  we obtain
\begin{eqnarray}
-i \epsilon  \Phi_{0}^{cart} -
 {1 \over \cos z} ({\partial \over \partial r} +{ \cos r \over  \sin r}) \Phi^{cart}_{1} -
 \nonumber
\\
-
{im \over \cos z \sin r}  \Phi^{cart}_{2} - ( {\partial \over
\partial z} - 2 \tan z) \Phi^{cart}_{3} =0 \; . \label{3.13c}
\end{eqnarray}

To use this relation in the above equations, we should transform (\ref{3.13c}) to cyclic basis:
\begin{eqnarray}
\Phi_{0} = \Phi^{cart}_{0}\; , \qquad \Phi_{2} = \Phi^{cart}_{3}\;
,
\nonumber
\\
\Phi_{3} - \Phi_{1} =  \sqrt{2}  \Phi_{1}^{cart}   \;,
 \qquad \Phi_{3} + \Phi_{1} =\sqrt{2}i\;  \Phi_{2}^{cart}\;;
\nonumber
\end{eqnarray}

\noindent thus we have
\begin{eqnarray}
-i \epsilon  \Phi_{0} -
 {1 \over \cos z} ({\partial \over \partial r} +{ \cos r \over  \sin r}){ \Phi_{3} - \Phi_{1} \over \sqrt{2}}
 -
\nonumber
\\
- {im \over \cos z \sin r}  { \Phi_{3} + \Phi_{1} \over \sqrt{2} i
} - ( {\partial \over \partial z} - 2 \tan z) \Phi_{2} =0 \; ,
\label{4.13d}
\end{eqnarray}

\noindent  that is
\begin{eqnarray}
-i\epsilon \Phi_{0} - {1 \over \cos z} \; b _{-} \; \Phi_{1} - {1
\over  \cos z} \; a _{+} \;\Phi_{3} - ( {\partial \over \partial
z} - 2 \tan z) \Phi_{2} =0 \; . \label{4.13e}
\end{eqnarray}

Now, let us turn to eqs.  (\ref{4.12a}) -- (\ref{4.12d}). First, let us consider the case
\underline{$\sigma \neq 0$}, when one must accept from the very beginning  such a restriction
$\Phi_{0}=0$;  correspondingly, equation become more simple
\begin{eqnarray}
\sigma \neq 0 \;, \qquad \Phi_{0}=0\;,
\nonumber
\\
  -b_{-}  \; E_{1}   - a_{+}  \;  E_{3}
-   \cos z ( {\partial  \over
\partial z } -2 \tan z) E_{2}
 = 0 \; ,
\label{4.14a}
\end{eqnarray}
\begin{eqnarray}
        i \epsilon  \;  E_{1}
 + i  \sigma\,  H_{1}  = M  \;  \Phi _{1}\;,
\nonumber
\\
  i \sigma\;  H_{2}
    + i \epsilon \;    E_{2}   = M  \; \Phi_{2}\;,
\nonumber
\\
  i \epsilon  \;  E_{3}
+ i \sigma \;  H_{3}   = M    \; \Phi_{3}\; , \label{4.14b}
\end{eqnarray}
\begin{eqnarray}
    - i  \epsilon  \;   \Phi_{1} = M \;      E_{1}\;,
\nonumber
\\
-i \epsilon    \Phi_{2}    = M  \;   E_{2} \;,
\nonumber
\\
  -i \epsilon  \;\Phi _{3}  = M    \;  E_{3}\;,
\label{4.14c}
\end{eqnarray}
\begin{eqnarray}
-   \sigma\,  \Phi_{1}   = M  \; H_{1}\;,
\nonumber
\\
  - i \sigma \;  \Phi_{2}    = M  \;   H_{2} \; ,
\nonumber
\\
 -i\sigma \;    \Phi_{3}   =  M  \;    H_{3}\;.
\label{4.14d}
\end{eqnarray}

Note that substituting (\ref{4.13d}) into  (\ref{4.13a}), one gets
\begin{eqnarray}
  -b_{-}  \; \Phi_{1}   - a_{+}  \;  \Phi_{3}
-   \cos z ( {\partial  \over
\partial z } -2 \tan z) \Phi_{2}
 = 0 \; ,
\label{4.14a'}
\end{eqnarray}

\noindent which coincides with the Lorentz condition (\ref{4.13e})  when $\Phi_{0}=0$.
Condition   (\ref{4.14a'}) can be simplified by the following substitutions
\begin{eqnarray}
\Phi_{1} = {\varphi_{1} \over \cos z} \; , \qquad  \Phi_{3} =
{\varphi_{3} \over \cos z} \; , \qquad \Phi_{2} = {1 \over
\cos^{2}z} \varphi_{2} \; ,
\nonumber
\end{eqnarray}

\noindent which results in
\begin{eqnarray}
  -b_{-}   \varphi_{1}   - a_{+}    \varphi_{3}
-    {\partial  \over
\partial z }  \varphi_{2}
 = 0 \; ;
\label{4.14a''}
\end{eqnarray}

\noindent in new variables  $b_{-}   \varphi_{1}  = \bar{\varphi}_{1} ,
\; a_{+}    \varphi_{3}= \bar{\varphi}_{3}$ it becomes yet simpler
\begin{eqnarray}
   \bar{\varphi}_{1}   + \bar{\varphi}_{3}+     {\partial  \over \partial z }  \varphi_{2}
 = 0 \; .
\label{4.14a'''}
\end{eqnarray}

\noindent
Remaining algebraic relations   will fixe  values of $\sigma$ and relative coefficients
 of various components
 \begin{eqnarray}
 \sigma = \pm i \sqrt{ \epsilon^{2} - M^{2} } \;  , \qquad  \Phi_{0}= 0\; ,
\nonumber
\\
  H_{j}  = - i{ \sigma \over M}  \;  \Phi_{j}\;, \qquad   E_{j} = { i \epsilon  \over M}  \Phi_{j}
 \; .
\label{4.15a}
\end{eqnarray}

Explicit  form of the main functions  $\Phi_{j}$ will be found below when exploring
helicity operator equations.

In \underline{masless case}  instead of  (\ref{4.15a}) we have
\begin{eqnarray}
 \sigma = \pm i  \epsilon  \;  , \qquad  \Phi_{0}= 0\; ,
\nonumber
\\
  H_{j}  = - i \sigma   \;  \Phi_{j}\;, \qquad   E_{j} =  i \epsilon  \; \Phi_{j}  \; .
\label{4.15b}
\end{eqnarray}

Now, let us consider \underline{the case   $\sigma=0$}, when the system  (\ref{4.12a}) -- (\ref{4.12d})
is
\begin{eqnarray}
  -b_{-}  \; E_{1}   - a_{+}  \;  E_{3}
-   \cos z ( {\partial  \over
\partial z } -2 \tan z) E_{2}
 = M   \; \cos z \Phi_{0} \; ,
\label{4.16a}
\end{eqnarray}
\begin{eqnarray}
        i \epsilon  \;  E_{1}
  = M  \;  \Phi _{1}\;,\qquad
       i \epsilon \;    E_{2}   = M  \; \Phi_{2}\;, \qquad
  i \epsilon  \;  E_{3}
   = M    \; \Phi_{3}\; ,
\label{4.16b}
\end{eqnarray}
\begin{eqnarray}
    a \; \Phi_{0}  - i  \epsilon  \; \cos z  \Phi_{1} = M \; \cos z     E_{1}\;,
\nonumber
\\
-i \epsilon    \Phi_{2}   - {\partial  \over \partial z}  \Phi
_{0} = M  \;   E_{2} \;,
\nonumber
\\
    b \;
\Phi_{0}  -i \epsilon  \;\cos z \Phi _{3}  = M    \; \cos z
E_{3}\;, \label{4.16c}
\end{eqnarray}
\begin{eqnarray}
0= M  \; H_{1}\;, \qquad
 0   = M  \;   H_{2} \; , \qquad
 0   =  M  \;    H_{3}\;.
\label{4.16d}
\end{eqnarray}

Note that allowing for (\ref{4.16b}), from  (\ref{4.16a}) it follows
\begin{eqnarray}
  -b_{-}  \; \Phi_{1}   - a_{+}  \;  \Phi_{3}
-   \cos z ( {\partial  \over
\partial z } -2 \tan z) \Phi_{2}
 = i \epsilon    \; \cos z \Phi_{0} \; ,
\label{4.16a'}
\end{eqnarray}

\noindent which coincides with the Lorentz condition.

Let us introduce  substitutions
\begin{eqnarray}
\Phi_{1} = {\varphi_{1} \over \cos z} \; , \qquad  \Phi_{3} =
{\varphi_{3} \over \cos z} \; , \qquad \Phi_{2} = {1 \over
\cos^{2}z} \varphi_{2} \; ,
\nonumber
\\
E_{1} = { e_{1} \over \cos z} \; , \qquad  E_{3} = {e_{3} \over
\cos z} \; , \qquad E_{2} = {1 \over \cos^{2}z} e_{2} \; ,
\nonumber
\\
b_{-} \varphi_{1} = \bar{\varphi}_{1} , \qquad a_{+} \varphi_{3} =
\bar{\varphi}_{3}\; , \qquad b_{-} e_{1} = \bar{e}_{1} , \qquad
a_{+} e_{3} = \bar{e}_{3}\; ,
\nonumber
\end{eqnarray}

\noindent then eqs. (\ref{4.16a}) -- (\ref{4.16d}) read
\begin{eqnarray}
   \bar{\varphi}_{1}   + \bar{\varphi}_{3}
+    {\partial  \over
\partial z }  \varphi_{2}
 = -  i \epsilon    \; \cos ^{2}z \Phi_{0} \; ,
\label{4.17a}
\end{eqnarray}
\begin{eqnarray}
        i \epsilon  \;  e_{1}
  = M  \;  \varphi _{1}\;, \qquad
       i \epsilon \;    e_{2}   = M  \; \varphi_{2}\;, \qquad
  i \epsilon  \;  e_{3}
   = M    \; \varphi_{3}\; ,
\label{4.17b}
\end{eqnarray}
\begin{eqnarray}
   \Delta  \; \Phi_{0}  - i  \epsilon  \;  \bar{\varphi}_{1} = M \; \bar{e}_{1}\;,
\nonumber
\\
-i \epsilon    \varphi_{2}   - \cos^{2}z {\partial  \over \partial
z}  \Phi _{0} = M  \;   e_{2} \;,
\nonumber
\\
   \Delta  \;
\Phi_{0}  -i \epsilon  \;\bar{\varphi} _{3}  = M    \;
\bar{e}_{3}\;, \label{4.17c}
\end{eqnarray}
\begin{eqnarray}
0=  H_{1}\;,\qquad
 0   =   H_{2} \; , \qquad
 0   =      H_{3}\;,
\label{4.17d}
\end{eqnarray}

\noindent  one should take into consideration identity $ \Delta = b_{-}a = a_{+}b$.

Below we will show from helicity operator eigenvalue equation that when
 $\sigma=0$ there must hold the following relationships
 \begin{eqnarray}
\bar{\varphi} _{1} = \bar{\varphi} _{3}  = \bar{\varphi}\;, \qquad
\bar{e} _{1} = \bar{e} _{3}  = \bar{e}\;, \qquad \bar{h} _{1} =
\bar{h} _{3}  = \bar{h}\;,
\nonumber
\\
\Delta \varphi_{2} = - \cos^{2} z {\partial \over  \partial z}
\bar{\varphi} \; ,\;
 \Delta e_{2} = - \cos^{2} z{\partial
\over  \partial z} \bar{e} \; ,\;
\Delta h_{2} = - \cos^{2}
z{\partial \over  \partial z} \bar{h} \; ; \label{4.18}
\end{eqnarray}

\noindent so that from  (\ref{4.17a}) -- (\ref{4.17d})   we get
\begin{eqnarray}
  - 2\bar{\varphi}-    {\partial  \over
\partial z }  \varphi_{2}
 = i \epsilon    \; \cos ^{2}z \; \Phi_{0} \; ,
\label{4.19a}
\end{eqnarray}
\begin{eqnarray}
       \bar{e}
  = { M \over  i \epsilon }  \;  \bar{\varphi}\;,\qquad
         e_{2}   = { M \over  i \epsilon }    \; \varphi_{2}\;,\qquad H_{j} = 0 \; ,
\label{4.19b}
\end{eqnarray}
\begin{eqnarray}
 ( \epsilon ^{2}-M^{2})    \varphi_{2}   - i\epsilon
\cos^{2} z {\partial  \over \partial z}  \Phi _{0} = 0  \;,
\nonumber
\\
  i\epsilon   \Delta  \; \Phi_{0}  + (\epsilon^{2} -M^{2})  \bar{\varphi} = 0\;.
\label{4.19c}
\end{eqnarray}

Acting on the first equation in  (\ref{4.19c}) by the operator $\partial_{z}$,
and excluding in second equation in (\ref{4.19c})  the variable
$\bar{\varphi}$  with the help of (\ref{4.19a}) -- thus we get
 \begin{eqnarray}
 ( \epsilon ^{2}-M^{2})  {\partial \over \partial z}   \varphi_{2}   - i\epsilon
{\partial \over \partial z} \cos^{2} z {\partial  \over \partial
z}  \Phi _{0} = 0  \;,
\nonumber
\\
 2 i\epsilon   \Delta  \; \Phi_{0}  - (\epsilon^{2} -M^{2})
( {\partial  \over
\partial z }  \varphi_{2}
 + i \epsilon    \; \cos ^{2}z \; \Phi_{0} )
 = 0\;.
 \nonumber
\end{eqnarray}

\noindent Summing these two o equations, we arrive at a second order equation
for $\Phi_{0}$
\begin{eqnarray}
- i\epsilon {\partial \over \partial z} \cos^{2} z {\partial
\over \partial z}  \Phi _{0} + 2 i\epsilon   \Delta  \; \Phi_{0}
- (\epsilon^{2} -M^{2})
 i \epsilon    \; \cos ^{2}z \; \Phi_{0}  = 0\;,
\nonumber
\end{eqnarray}

\noindent that is
\begin{eqnarray}
\left ( -2    \Delta + {\partial \over \partial z} \cos^{2} z
{\partial  \over \partial z}
 + (\epsilon^{2} -M^{2})      \; \cos ^{2}z \right ) \Phi_{0}  = 0\; .
\label{4.20a}\end{eqnarray}

\noindent In eq.  (\ref{4.20a}), the variables are separated straightforwardly

\begin{eqnarray}
\Phi_{0} (r,z) = \Phi_{0} (r)  \Phi_{0} (z)\;, \qquad
{1 \over \Phi_{0} (r) }  \; ( 2\Delta  ) \Phi_{0} (r)  = \Lambda
\; ,
\nonumber
\\
{1 \over \Phi_{0} (z) } \left ( {d \over d z} \cos^{2} z {d  \over
d z}
 + (\epsilon^{2} -M^{2})      \; \cos ^{2}z \right ) \Phi_{0}(z)  = \Lambda\; .
\label{4.20b}
\end{eqnarray}

In the same manner, with the help of (\ref{4.19a}) on can exclude the function $\Phi_{0}$  from
second equation in (\ref{4.19c})
\begin{eqnarray}
\Delta(    - 2\bar{\varphi}-    {\partial  \over
\partial z }  \varphi_{2} )  + (\epsilon^{2} -M^{2})  \cos^{2} z    \bar{\varphi} = 0\;,
\nonumber
\end{eqnarray}

\noindent and further excluding the variable  $\Delta \varphi_{2}$ with the help of (\ref{4.18}) we arrive at
a second order equation for $\bar{\varphi} $
\begin{eqnarray}
  \left (   - 2\Delta  +    {\partial  \over
\partial z }  \cos^{2} z { \partial \over  \partial z}
  + (\epsilon^{2} -M^{2})  \cos^{2} z    \right ) \bar{\varphi} = 0\;.
\label{4.21a}
\end{eqnarray}

\noindent In this equation, the variable are separated as well
\begin{eqnarray}
\bar{\varphi} (r,z) = \bar{\varphi} (r)  \bar{\varphi} (z)\;,
\nonumber
\\
{1 \over \bar{\varphi} (r) }  \; ( 2\Delta  ) \bar{\varphi}(r)  =
\Lambda \; ,
\nonumber
\\
{1 \over \bar{\varphi} (z) } \left ( {d \over d z} \cos^{2} z {d
\over d z}
 + (\epsilon^{2} -M^{2})      \; \cos ^{2}z \right ) \bar{\varphi} (z)  = \Lambda\; .
\label{4.21b}
\end{eqnarray}

\noindent Note, that from the first equation in (\ref{4.19c}) it follows  an expression for
$\varphi_{2}$
\begin{eqnarray}
     \varphi_{2}   =  { i\epsilon \; \cos^{2} z \over ( \epsilon ^{2}-M^{2})}\;
 {\partial  \over \partial z}  \Phi _{0}  \; .
\label{4.22}
\end{eqnarray}

One can easily verify consistency of the relations obtained. Indeed, let u s act on eq.
(\ref{4.22}) by the operator
 $\Delta$
\begin{eqnarray}
   \Delta  \varphi_{2}   =  { i\epsilon \over ( \epsilon ^{2}-M^{2})} \Delta
\cos^{2} z {\partial  \over \partial z}  \Phi _{0} = 0  \;.
\nonumber
\end{eqnarray}

\noindent Further, allowing for  (\ref{4.18}) we get
\begin{eqnarray}
   - \cos^{2} z {\partial \over \partial z} \bar{\varphi}
  =  { i\epsilon \over ( \epsilon ^{2}-M^{2})} \Delta
\cos^{2} z {\partial  \over \partial z}  \Phi _{0} = 0  \;,
\nonumber
\end{eqnarray}

\noindent from whence it follows
\begin{eqnarray}
  i\epsilon   \Delta  \; \Phi_{0}  + (\epsilon^{2} -M^{2})  \bar{\varphi} = 0\;,
\nonumber
\end{eqnarray}

\noindent which is an identity
\begin{eqnarray}
   -  {\partial \over \partial z} \bar{\varphi}
  \equiv  -  { 1 \over ( \epsilon ^{2}-M^{2})} (\epsilon^{2} -M^{2})
 {\partial  \over \partial z}     \bar{\varphi}   \; .
\nonumber
\end{eqnarray}

Now, let us turn to equations steaming from diagonalization of helicity operator.
In (\ref{4.11a}) -- (\ref{4.11d})  owe can notice three similar groups of equations.
For instance, equations for $H_{i}$ are
\begin{eqnarray}
 a\;    H_{2}  + \cos z ({ \partial  \over \partial  z} -\tan z )  H_{1} = \sigma\, \cos z H_{1} \;,
\nonumber
\\
- b_{-} H_{1} + a_{+} H_{3} = \sigma\; \cos z \;  H_{2}
  \; ,
\nonumber
\\
 -b\;    H_{2}    - \cos z ({ \partial  \over \partial  z} -\tan z )   H_{3}=  \sigma \;  \cos z  \; H_{3}\; .
\label{4.23a}
\end{eqnarray}

\noindent With the help op substitutions
\begin{eqnarray}
H_{1} = {1 \over \cos z} h_{1}(r,z) \;, \qquad  H_{2} = {1 \over
\cos^{2} z} h_{2}(r,z) \;, \qquad H_{3} = {1 \over \cos z}
h_{3}(r,z) \;, \label{4.23b}
\end{eqnarray}

\noindent they are simplified
\begin{eqnarray}
 a\;    h_{2}  = \cos^{2}z  (+ \sigma -  { \partial  \over \partial  z}  ) \;  h_{1} \;,
\nonumber
\\
- b_{-} h_{1} + a_{+} h_{3} = \sigma\;   h_{2}
  \; ,
  \nonumber
  \\
 b\;    h_{2}    = \cos^{2} z ( -\sigma -  { \partial  \over \partial  z}) \;   h_{3}\; .
\label{4.23c}
\end{eqnarray}

\noindent Let us introduce new variables
\begin{eqnarray}
b_{-}h_{1} = \bar{h}_{1} \;, \qquad a_{+}h_{3} = \bar{h}_{3} \; ;
\label{4.24a}
\end{eqnarray}

\noindent from   (\ref{4.23c}) it follows
\begin{eqnarray}
 b_{-} a\;    h_{2}    =\cos^{2} z ( \sigma -{ \partial  \over \partial  z})    \bar{h}_{1}    \;,
\nonumber
\\
\bar{h}_{3} - \bar{h}_{1}   = \sigma\;   h_{2}
  \; ,
\nonumber
\\
 a_{+} b\;    h_{2}         = \cos^{2} z (- \sigma -{ \partial  \over \partial  z})  \;  \bar{h}_{3}  \;.
\label{4.24b}
\end{eqnarray}

\noindent Note that first and third equations contain one the same second order operator
\begin{eqnarray}
b_{-}a =a_{+}b = {1 \over 2} \left ( -{\partial ^{2} \over
\partial r^{2}} - {\cos r \over \sin r} {\partial \over \partial
r} + {m^{2} \over \sin^{2} r} \right ) = \Delta \; . \label{4.24c}
\end{eqnarray}

First, let us consider the case  $\sigma \neq 0$. Equating the right-hand sides of the first and third equations
in  (\ref{4.24b}), we get
\begin{eqnarray}
 \sigma ( \bar{h}_{1}  + \bar{h}_{3} )   =
-{ \partial  \over \partial  z}   (\bar{h}_{3}  - \bar{h}_{1} ) =-
\sigma { \partial  \over \partial  z} h_{2} \; ; \label{4.25a}
\end{eqnarray}

\noindent that is
\begin{eqnarray}
  \bar{h}_{3}  + \bar{h}_{1}    =  - { \partial  \over \partial  z} h_{2}\;, \qquad
\bar{h}_{3}   - \bar{h}_{1}    = \sigma\;   h_{2}
  \; .
\nonumber
\end{eqnarray}

\noindent Thus, we arrive at
expression for  $\bar{h}_{1} $ and  $\bar{h}_{3} $ through $h_{2}$
\begin{eqnarray}
\bar{h}_{3}  = {1 \over 2}(+\sigma -{ \partial  \over \partial
z}) h_{2} \; , \qquad \bar{h}_{1}  = {1 \over 2}(-\sigma -{
\partial  \over \partial  z}) h_{2} \; . \label{4.25a}
\end{eqnarray}

\noindent In turn, substituting (\ref{4.25a}) into   (\ref{4.24b}) we  obtain one the same second order equation for
$h_{2}$
\begin{eqnarray}
 b_{-} a\;    h_{2}    =\cos^{2} z ( \sigma -{ \partial  \over \partial  z})
  {1 \over 2}(-\sigma -{ \partial  \over \partial  z}) h_{2}    \;,
\nonumber
\\
 a_{+} b\;    h_{2}
  = \cos^{2} z (- \sigma -{ \partial  \over \partial  z})  \;
   {1 \over 2}(+\sigma -{ \partial  \over \partial  z}) h_{2}
 \;.
\label{4.26a}
\end{eqnarray}

The variables in   (\ref{4.26a}) are separated straightforwardly
\begin{eqnarray}
h_{2}(r,z)  =h_{2}(r) \;  h_{2}(z)\;,
\nonumber
\\
{1 \over h_{2}(r)}  \; (2b_{-} a ) \;    h_{2}(r)    = {1 \over
h_{2}(z)} \cos^{2} z ( {d^{2}   \over d  z^{2} } - \sigma ^{2} )
   h_{2}  (z) = \Lambda  \;,
\nonumber
\end{eqnarray}

\noindent from whence it follows separated  differential equations
\begin{eqnarray}
 (2b_{-} a ) \;    h_{2} (r) = \Lambda \; h_{2} (r) \; ,
\label{4.26b}
\\
  ({d^{2}   \over d  z^{2} } - \sigma ^{2} )
   h_{2}(z)   =  {\Lambda \over \cos^{2} z } \;  h_{2} (z) \; .
\label{4.26c}
\end{eqnarray}

Similar results are valid for functions $e_{i}$ and $\varphi_{i}$:
\begin{eqnarray}
 (2b_{-} a ) \;    e_{2} (r) = \Lambda \; e_{2} (r) \; ,
\nonumber
\\
  ({d^{2}   \over d  z^{2} } - \sigma ^{2} )
   e_{2}(z)   =  {\Lambda \over \cos^{2} z } \;  e_{2} (z) \; ,
\nonumber
\\
\bar{e}_{1}  = {1 \over 2}(-\sigma -{ \partial  \over \partial
z}) e_{2} \; ,\qquad \bar{e}_{3}  = {1 \over 2}(+\sigma -{
\partial  \over \partial  z}) e_{2} \; ; \label{4.27}
\end{eqnarray}

\begin{eqnarray}
 (2b_{-} a ) \;    \varphi_{2} (r) = \Lambda \; \varphi_{2} (r) \; ,
\nonumber
\\
  ({d^{2}   \over d  z^{2} } - \sigma ^{2} )
   \varphi_{2}(z)   =  {\Lambda \over \cos^{2} z } \;  \varphi_{2} (z) \; ,
\nonumber
\\
\bar{\varphi}_{1}  = {1 \over 2}(-\sigma -{ \partial  \over
\partial  z}) \varphi_{2} \; , \qquad \bar{\varphi}_{3}  = {1
\over 2}(+\sigma -{ \partial  \over \partial  z}) \varphi_{2} \; .
\label{4.28}
\end{eqnarray}

Now, let us turn to
the system  (\ref{4.24b}) when $\sigma =0$; it gives
\begin{eqnarray}
\bar{h}_{3} = \bar{h}_{1} = \bar{h}   \; ,
\nonumber
\\
 b_{-} a\;    h_{2}    = -\cos^{2} z { \partial  \over \partial  z}    \bar{h}    \;,
\nonumber
\\
 a_{+} b\;    h_{2}         = -  \cos^{2} z { \partial  \over \partial  z}  \;  \bar{h} \;.
\label{4.29}
\end{eqnarray}

\noindent Just these relations were used above starting with (\ref{4.18}).

Let us construct solutions of eqs. (\ref{4.28}):
\begin{eqnarray}
 (2b_{-} a ) \;    \varphi_{2} (r) = \Lambda \; \varphi_{2} (r) \; ,
\nonumber
\\
  ({d^{2}   \over d  z^{2} } - \sigma ^{2} )
   \varphi_{2}(z)   =  {\Lambda \over \cos^{2} z } \;  \varphi_{2} (z) \; ,
\nonumber
\\
\bar{\varphi}_{1}  = {1 \over 2}(-\sigma -{ \partial  \over
\partial  z}) \varphi_{2} \; , \qquad \bar{\varphi}_{3}  = {1
\over 2}(+\sigma -{ \partial  \over \partial  z}) \varphi_{2} \; .
\label{A.1}
\end{eqnarray}

In the radial equation
\begin{eqnarray}
 (2b_{-} a ) \;    \varphi_{2} (r) = \Lambda \; \varphi_{2} (r) \; ,
\nonumber
\end{eqnarray}

\noindent or
\begin{eqnarray}
 \left (
{d ^{2} \over d r^{2}} + {\cos r \over \sin r} {d \over d r}
-{m^{2} \over \sin^{2} r}  + \Lambda \right )\;    \varphi_{2} (r)
=0\; ; \label{A.2a}
\end{eqnarray}

\noindent let us introduce a new variable
 $ 1-\cos r=2\,x\,, \;x\in
[0\,,\;1] $:
\begin{eqnarray}
x\,(1-x)\,{d^{2}\varphi_{2}\over dx^{2}}+(1-2\,x)\,{d
\varphi_{2}\over dx}+\left(\Lambda-{1\over 4}\,{m^{2}\over
x}-{1\over 4}\,{m^{2}\over 1-x}\right)\,\varphi_{2}=0\,
\label{A.2b}
\end{eqnarray}

\noindent and make a substitution
$\varphi_{2}=x^{a}\,(1-x)^{b}\,F_{2}$ ; thus we arrive at
\begin{eqnarray}
x\,(1-x)\,{d^{2}F_{2}\over
dx^{2}}+\left[2\,a+1-(2\,a+2\,b+2)\,x\right]\,{dF_{2}\over dx}\,+
\nonumber
\\
+\left[-(a+b)\,(a+b+1)+\Lambda+{1\over 4}\,{4\,a^{2}-m^{2}\over
x}+{1\over 4}\,{4\,b^{2}-m^{2}\over 1-x}\right]\,F_{2}=0\,.
\label{A.2c}
\end{eqnarray}

\noindent At  $a\,,\;b$ taken according to
\begin{eqnarray}
a=\pm{\mid m \mid \over 2}\,,\qquad b=\pm{\mid m \mid \over 2}\,,
\label{A.2d}
\end{eqnarray}

\noindent  eq. (\ref{A.2c}) becomes simpler
\begin{eqnarray}
x\,(1-x)\,{d^{2}F_{2}\over
dx^{2}}+\left[2\,a+1-(2\,a+2\,b+2)\,x\right]\,{dF_{2}\over dx}\,-
\nonumber
\\
-\left[(a+b)\,(a+b+1)-\Lambda\right]\,F_{2}=0\, \label{A.3a}
\end{eqnarray}

\noindent it represents a hypergeometric equations  \cite{Bateman-Erdei-1973} with parameters
\begin{eqnarray}
\alpha=a+b+{1\over 2}-{1\over 2}\,\sqrt{1+4\Lambda}\,,
\nonumber
\\
\beta=a+b+{1\over 2}+{1\over 2}\,\sqrt{1+4\Lambda}\,,\qquad
\gamma=2\,a+1\,. \label{A.3b}
\end{eqnarray}

By physical reason for $a,\,b$  we take positive values
\begin{eqnarray}
a=+{\mid m \mid \over 2}\,,\qquad b=+{\mid m \mid \over 2}\,;
\label{A.4a}
\end{eqnarray}

\noindent so the radial function looks
as
\begin{eqnarray}
\varphi_{2} (r) = \left ( \sin {r\over 2} \right )^{+\mid m \mid }
\left ( \cos {r\over 2} \right )^{+\mid m \mid }  F(\alpha, \beta,
\gamma; \sin^{2} {r \over 2}) \; ; \label{A.4b}
\end{eqnarray}

\noindent
 these solutions vanish at the points $r = 0, + \pi$.
To have polynomials one should  impose the known condition
$\alpha=-n_{r} $, so we get a quantization rule
\begin{eqnarray}
+{   \sqrt{1 + 4 \Lambda} \over 2}  = n_{r} + \mid m \mid + {1
\over 2}\;; \label{A.4c}
\end{eqnarray}

\noindent corresponding solutions are defined according to
\begin{eqnarray}
\varphi_{2}=\left(\sin{r\over 2}\right)^{+\mid m \mid}\,
\left(\cos {r\over 2}\right)^{+\mid m \mid} \times
\nonumber
\\
\times
F (-n,\;2\,\mid m \mid
+1+n,\;\mid m \mid +1;\;-\sin^{2}{r\over 2} )\,. \label{A.4d}
\end{eqnarray}

Now, let us solve equation  (\ref{A.1}) in variable  $z$
\begin{eqnarray}
  ({d^{2}   \over d  z^{2} } - \sigma ^{2} )
   \varphi_{2}(z)   =  {\Lambda \over \cos^{2} z } \;  \varphi_{2}
   (z), \qquad  -\sigma^{2} = \epsilon^{2} - M^{2} \; .
\label{A.5}
\end{eqnarray}

\noindent A first step is to introduce a new variable  (which distinguish between conjugated point
$+z$  and  $-z$ of spherical space)
\begin{eqnarray}
y = {1 + i \tan z \over 2}\; , \qquad 1- y = {1 - i \tan z
\over2}\; ; \label{A.6a}
\end{eqnarray}

\noindent if  $z \in [-\pi /2 , + \pi /2 ]$, the variable  $y$
belong s to a vertical line in the complex plane
\begin{eqnarray}
y =(  {1 \over 2} - i\infty,  {1 \over 2} + i\infty  )\,. \label{A.6b}
\end{eqnarray}

Allowing for
\begin{eqnarray}
{d \over dz} ={i\over 2} {1 \over \cos^{2} z}  {d \over dy}=
 2i y (1-y) {d \over dy}\; ,
\nonumber
\\
{\Lambda \over \cos^{2} z } = 4\Lambda y(1-y) \;, \label{A.6c}
\end{eqnarray}

\noindent eq. (\ref{A.5}) reduces to
\begin{eqnarray}
\left ( y(1-y) {d^{2} \over dy^{2} } + (1-2y) {d \over dy} +
\Lambda - {\epsilon^{2} - M^{2} \over 4 y(1-y)} \right
)\varphi_{2} = 0 \; . \label{A.7}
\end{eqnarray}

In the region  $y\sim 0$, eq.  (\ref{A.7}) becomes simpler
\begin{eqnarray}
\left ( y {d^{2} \over dy^{2} } + {d \over dy}
 - {\epsilon^{2} - M^{2} \over 4 y} \right )\varphi_{2} = 0\;, \qquad  \varphi_{2} \sim  y^{a} \; ,
\nonumber
\\
a(a-1) +a -{\epsilon^{2} - M^{2} \over 4}  = 0\,, \qquad a = \pm
{\sqrt{\epsilon^{2} - M^{2} } \over 2} \; . \label{A.8a}
\end{eqnarray}

In the region  $y\sim 1$, eq.  (\ref{A.7}) becomes simpler as well
\begin{eqnarray}
\left ( (1-y) {d^{2} \over dy^{2} } - {d \over dy}
 - {\epsilon^{2} - M^{2} \over 4 (1-y)} \right )\varphi_{2} = 0\;, \qquad
 \varphi_{2} \sim (1-y)^{b} \; ,
\nonumber
\\
b(b-1) +b -{\epsilon^{2} - M^{2} \over 4}  = 0\,, \qquad b = \pm
{\sqrt{\epsilon^{2} - M^{2} } \over 2} \; . \label{A.8b}
\end{eqnarray}

Searching solutions in the form
$
\varphi_{2}  (y)  = y^{a} (1-y)^{b} F (y)
$
for  $F (y)$ we have



\begin{eqnarray}
y(1-y)F'' + F' [ (2a+1) - y(2a+2b +2)] F'+
\nonumber
\\
+\left [-(a+b)(a+b+1)  + \Lambda + {1\over y} \left (a^{2} -
{\epsilon^{2}-M^{2} \over 4} \right )+ \right.
\nonumber
\\
\left. +
 {1\over 1- y} \left (a^{2} - {\epsilon^{2}-M^{2} \over 4} \right ) \right ]
F = 0 \; .
\label{A.9}
\end{eqnarray}

\noindent Let it be
\begin{eqnarray}
a = \pm {\sqrt{\epsilon^{2}-M^{2}} \over 2}, \qquad b = \pm
{\sqrt{\epsilon^{2}-M^{2}} \over 2}, \qquad
\nonumber
\\
\varphi_{2} = \left ( {1 + i \tan z \over 2} \right )^{a}  \left (
{1 - i \tan z \over 2} \right )^{b} F \; ; \label{A.10}
\end{eqnarray}

\noindent there are four possibilities depending on $a,b$
\begin{eqnarray}
a=b = - {\sqrt{\epsilon^{2}-M^{2}} \over 2}, \qquad \varphi_{2}
\sim  \cos^{-2a} z \;  F  (z) \; ;
\nonumber
\\
a=b = + {\sqrt{\epsilon^{2}-M^{2}} \over 2}, \qquad \varphi_{2}
\sim  \cos^{-2a} z \;  F  (z) \; ;
\nonumber
\\
a=-b \; , \qquad \varphi_{2}  \sim   + e^{+2ia z}  \;  F  (z) \; ;
\nonumber
\\
a=-b  \; , \qquad \varphi_{2}  \sim   - e^{-2ia z}  \;  F  (z) \;
. \label{A.11}
\end{eqnarray}

As relations  (\ref{A.10}) hold, eq.   (\ref{A.9}) takes the form
\begin{eqnarray}
y(1-y)F'' + F' [ (2a+1) - y(2a+2b +2)] F' -
\nonumber
\\
- [(a+b)(a+b+1)  -  \Lambda   \; ] \; F = 0 \; , \label{A.12a}
\end{eqnarray}

\noindent which can be recognized as a hypergeometric equation \cite{Bateman-Erdei-1973}
\begin{eqnarray}
y(1-y)\;  F + [ \; \gamma - (\alpha + \beta +1) y \; ]\;  F' -
\alpha \beta \; F = 0 \; , \label{A.12b}
\\
\gamma = (2a+1)\; ,\qquad
\alpha = a + b +{1\over 2} + {\sqrt{4\Lambda +1 }\over 2}  \; ,
\nonumber
\\
 \beta = a + b +{1\over 2}
- {\sqrt{4\Lambda +1 }\over 2}  \; . \label{A.12c}
\end{eqnarray}

In this point we should notice that the spectrum for $\Lambda$ has been found
 (see  (\ref{A.4c})) from analyzing the differential equation in the variable  $r$,
 therefore  now to produce a spectrum for energy one must consider
 the cases  with  $a=b$.

There arise two possibilities.

The first:
\begin{eqnarray}
2a=2b=   \sqrt{\epsilon^{2}-M^{2}} \; , \qquad \beta = -n_{z}\,,
\nonumber
\\
-\sqrt{\epsilon^{2} - M^{2}}  =   n_{z} +{1 \over 2} - {
\sqrt{4\Lambda +1 }\over 2}< 0 \; ,
\nonumber
\\
\varphi_{2}  \sim  ( \cos z ) ^{- \sqrt{\epsilon^{2}-M^{2}} }  \;
P_{n}  ({e^{iz} \over 2 \cos z})\;
 , \qquad y = {1 + i \tan z \over 2} = {e^{iz} \over 2 \cos z}\,,
\label{A.13}
\end{eqnarray}

\noindent because those solutions tends to  infinity at  $z = \pm
\pi $ they cannot describe physical bound states.

The second:
\begin{eqnarray}
-2a=-2b= +  \sqrt{\epsilon^{2}-M^{2}} \; , \qquad \alpha = -n_{z}\,,
\nonumber
\\
+\sqrt{\epsilon^{2} - M^{2}}  =  n_{z} + {1 \over 2} +
{\sqrt{4\Lambda +1 }\over 2}> 0 \; ,
\nonumber
\\
\varphi_{2}  \sim  ( \cos z ) ^{+ \sqrt{\epsilon^{2}-M^{2}} }  \;
P_{n}  ({e^{iz} \over 2 \cos z})\;
 , \qquad y = {1 + i \tan z \over 2} = {e^{iz} \over 2 \cos z}\,.
\label{A.14}
\end{eqnarray}

These solutions  are finite at the points  $z = \pm \pi
/2 $ and they describe bound states.

In the formula for $\sqrt{\epsilon^{2} - M^{2}}$  (\ref{A.14}) one must take into account
the quantization rule  $\Lambda $ in (\ref{A.4c}) -- thus we arrive
at the formula  determining values of energy by two discrete  quantum numbers.
\begin{eqnarray}
+\sqrt{\epsilon^{2} - M^{2}}  =  n_{z} +  n_{r} + \mid m \mid + 1
\; ; \label{A.15}
\end{eqnarray}

\noindent remember that these formulas concern
the non-zero values for helicity operator
 $\sigma =  \pm i
\sqrt{\epsilon^{2} - M^{2} }$.

It remains  to specify  the energy spectrum
for states with
 $\sigma=0$ which are determined by the equations
 \begin{eqnarray}
  \left (   - 2\Delta  +    {\partial  \over
\partial z }  \cos^{2} z { \partial \over  \partial z}
  + (\epsilon^{2} -M^{2})  \cos^{2} z    \right ) \bar{\varphi} = 0\;,
  \nonumber
\\
\bar{\varphi} (r,z) = \bar{\varphi} (r)  \bar{\varphi} (z)\;,
\nonumber
\\
{1 \over \bar{\varphi} (r) }  \; ( 2\Delta  ) \bar{\varphi}(r)  =
\Lambda \; ,
\nonumber
\\
{1 \over \bar{\varphi} (z) } \left ( {d \over d
z} \cos^{2} z {d   \over d z}
 + (\epsilon^{2} -M^{2})      \; \cos ^{2}z \right ) \bar{\varphi} (z)  = \Lambda\; .
\label{A.16}
\end{eqnarray}

Equation in the variable  $r$ has been solved above.
The equation in  $z $ variable
\begin{eqnarray}
\left ( {d^{2} \over dz^{2}}  -2 {\sin z \over \cos z} {d \over
dz} + \epsilon^{2} - M^{2}- {\Lambda \over \cos^{2} z} \right )
\varphi (z) = 0
\nonumber
\end{eqnarray}

\noindent with the use of substitution
$
\varphi = {1 \over \cos z}  f (z)
$
 reduces to
 \begin{eqnarray}
 {d^{2}f \over dz^{2}}   + \left(\epsilon^{2} - M^{2}+1-
{\Lambda \over \cos^{2} z} \right ) f (z) = 0\,.\label{A.18}
\end{eqnarray}

\noindent It coincides with
(\ref{A.5})
\begin{eqnarray}
  {d^{2}  \varphi_{2}  \over d  z^{2} }+\left(\epsilon^{2} - M^{2}
     -  {\Lambda \over \cos^{2} z }\right) \;  \varphi_{2}
   (z)=0\,,
\nonumber
\end{eqnarray}

\noindent with one formal change
\begin{eqnarray}
\epsilon^{2} - M^{2}\;\;\rightarrow\;\;\epsilon^{2} - M^{2}+1\,.
\nonumber
\end{eqnarray}

\noindent Therefore, solutions of (\ref{A.18}) are written straightforwardly
\begin{eqnarray}
f = \left ( {1 + i \tan z \over 2} \right )^{a}  \left ( {1 - i \tan
z \over 2} \right )^{b} F \,\left(\alpha,\;\beta,\;\gamma;\;{1 + i
\tan z \over 2}\right)\; ,
\nonumber
\end{eqnarray}

\noindent where  $F$ stand for a hypergeometric function  \cite{Bateman-Erdei-1973} with parameters
\begin{eqnarray}
\alpha = a + b +{1\over 2} + {\sqrt{4\Lambda +1 }\over 2}  \; ,
\nonumber
\\
 \beta = a + b +{1\over 2}
- {\sqrt{4\Lambda +1 }\over 2}  \; ,
\gamma = (2a+1)\; .
\nonumber
\end{eqnarray}

\noindent To bound states correspond  $a,\;b$ defined as
\begin{eqnarray}
a =b= - {\sqrt{\epsilon^{2}-M^{2}+1} \over 2}\,.
\nonumber
\end{eqnarray}

The quantization rule
 $\alpha=-n_{z}$ gives
\begin{eqnarray}
+\sqrt{\epsilon^{2} - M^{2}+1}  =  n_{z} + {1 \over 2} +
{\sqrt{4\Lambda +1 }\over 2}> 0 \; .
\nonumber
\end{eqnarray}

\noindent Thus, allowing for quantization for  $\Lambda $
 (\ref{A.4c}) we get a formulas for energy levels
 \begin{eqnarray}
+\sqrt{\epsilon^{2} - M^{2}+1}  =  n_{z} +  n_{r} + \mid m \mid +
1 \; ;
\end{eqnarray}

\noindent  it refers to the case
of
$\sigma=0$.

{\bf  Conclusion:}

Let us summarize result.

Spin 1  particle is investigated in  3-dimensional curved
space of constant positive curvature. An extended helicity
operator is defined and the variables are separated in a
tetrad-based 10-dimensional Duffin-Kemmer equation in quasi
cylindrical coordinates. The problem is solved exactly in
hypergeometric functions, the energy spectrum determined by three
discrete quantum numbers is obtained. Transition to a massless
case of electromagnetic field is performed.

The given problem  can represent some interest as an exactly solvable model for
describing  composite systems  (particles) of spin 1  or electromagnetic fields
in the non-trivial space-time background, modeling
the presence of a finite  3-dimensional box.

\end{document}